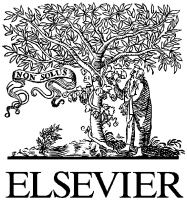

# Analysis of Maximum Likelihood and Mahalanobis Distance for Identifying Cheating Anchor Nodes

Jeril Kuriakose[a], Amruth V[b], Sandesh A G[c], Jampu Venkata Naveenbabu[a], Mohammed Shahid[b], and Ashish Shetty[b]

[a] School of Computing and Information Technology (SCIT), Manipal University Jaipur, Jaipur, India.
Email: jeril@muj.manipal.edu

[b] Department of Information Science and Engineering, Bearys Institute of Technology, Mangalore, India.

[c] Associate Consultant, Einsys Consulting Pvt. Ltd., Bangalore, India.

**Abstract**

Malicious anchor nodes will constantly hinder genuine and appropriate localization. Discovering the malicious or vulnerable anchor node is an essential problem in wireless sensor networks (WSNs). In wireless sensor networks, anchor nodes are the nodes that know its current location. Neighboring nodes or non-anchor nodes calculate its location (or its location reference) with the help of anchor nodes. Ingenuous localization is not possible in the presence of a cheating anchor node or a cheating node. Nowadays, it's a challenging task to identify the cheating anchor node or cheating node in a network. Even after finding out the location of the cheating anchor node, there is no assurance, that the identified node is legitimate or not. This paper aims to localize the cheating anchor nodes using trilateration algorithm and later associate it with maximum likelihood expectation technique (MLE), and Mahalanobis distance to obtain maximum accuracy in identifying malicious or cheating anchor nodes during localization. We were able to attain a considerable reduction in the error achieved during localization. For implementation purpose we simulated our scheme using ns-3 network simulator.

## 1. Introduction

Wireless ad hoc and sensor networks are on a steady rise in the recent decade. This is because of their reduced cost in deployment and maintenance. Advancements in radio frequency spectrum also carved way for the improvement in the data rate for communication. Many devices belong to wireless ad hoc and sensor networks; one among them is anchor node [1 – 8]. Anchor nodes are the nodes that know its current location. Neighboring nodes or non-anchor nodes calculate its location (or location reference) with the help of anchor nodes, and its working is quite referable to Light House.

The location of the nodes plays a significant role in many areas as routing, surveillance and monitoring, military etc. The sensor nodes must know their location reference to carryout location-based routing (LR) [9 – 12]. To find out the shortest route, the location aided routing (LAR) [13 – 15] makes use of the locality reference of the sensor nodes. In some industries the sensor nodes are used to identify minute changes as pressure, temperature and gas leak, and in military, robots are used to detect landmines where in both the cases location information plays a key part.

Anchor nodes can also be used to find the current location of any device (mobile phones, objects and people). It does that by transmitting anchor frames periodically or at regular intervals. Usually anchor frames are used to advertise the occurrence of a wireless modem or an Access Point (AP). Each anchor frame carries some details about the configuration of AP and a little security information for the clients.

When the technologies are on a massive upswing, the need for security of the relevant technologies arises. There can be several occasion where the anchor nodes can be vulnerable to security breach. Because of the security breach the anchor node starts cheating by providing false information. In the presence of cheating anchor nodes the chances of localization drastically decreases. Many papers [16 – 19] discuss about the localization of cheating anchor nodes, but with inconsistent accuracy. So, to overcome this, we localize the cheating or vulnerable anchor node using trilateration technique, and associate the results with maximum likelihood expectation technique [20, 32] and Mahalanobis distance [34]. No such scheme has been used till now to identify the malicious anchor nodes.

*Jeril Kuriakose, et.al.*

**Organization of the paper** – Section 2 provides the localization using trilateration algorithm and section 3 and section 4 studies the maximum likelihood expectation and Mahalanobis distance, respectively. Simulation and results are covered in section 5, section 6 explicates few future events and section 7 concludes the paper.

## 2. Localization using Trilateration Algorithm

Anchor nodes are widely used for tacking and localization; whereas now-a-days it is also used for navigation and route-identification. With the help of anchor nodes, a user can find out his current location. Consider a scenario like a hotel or museum, there may be many occasions where people go out of track. This can be flabbergasted by installing anchor nodes installed in various locations, so that people can trace out there location very easily and it is possible only when the anchor nodes are authentic. Now-a-days hackers are on a rise; anybody can easily get into any system and change its settings. Similarly, they can hack any anchor nodes and change its location reference to some other false location reference, making people lose their track; thus leading to a bad imprint about the system (i.e., hotel, museum).

An attack is exemplified in fig. 1 and fig. 2. Fig. 1 shows the initial deployment of anchor nodes $A_1$, $A_2$, $A_3$; with location reference $(x_1, y_1)$, $(x_2, y_2)$, $(x_3, y_3)$; and distance $L_1$, $L_2$, $L_3$; respectively, from the trilateration point T, having location reference $(x_t, y_t)$. Fig. 2 demonstrates the logical deployment of anchor nodes after the attack i.e., multiple changes in location reference of anchor node $A_2$.

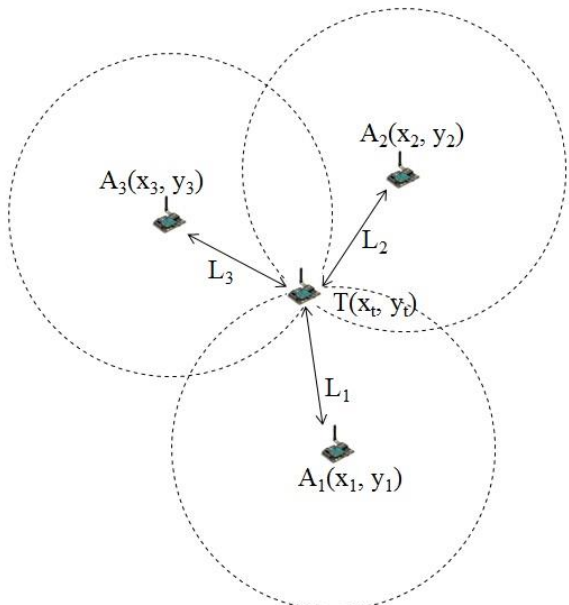
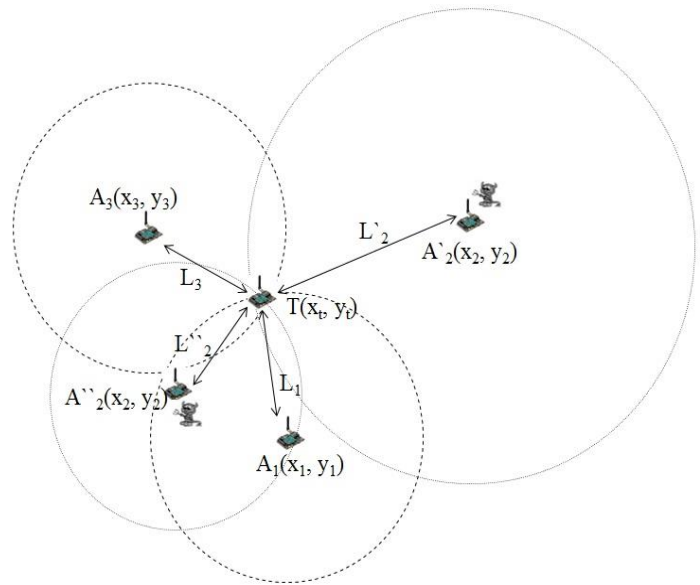

Fig. 1. Initial setup of anchor nodes.                    Fig. 2. Anchor nodes after attack.

The three dimensional location coordinate of any device or node can be estimated using trilateration calculations. Trilateration technique uses distance measurements rather than angular measurements; latter technique is also used in many localization techniques [21 – 23]. Using some iterative schemes like least square, least median square [17], least trimmed square [24] and gradient descent [25], can equitably increase the accuracy of trilateration technique.

Trilateration techniques use the distance measurement between the nodes to calculate the location reference. The distances between the nodes are identified using Received Signal Strength (RSSI) [26, 33] or Time of Arrival (ToA) [27, 28, 33] or Time Difference of Arrival (TDoA) [29, 30, 33]. When a node (requesting node) wants to identify its location information using trilateration technique, it does with the help of three or more neighboring anchor nodes. The exemplification of trilateration techniques is as follows:

a. A node that wants to find its location reference (or location coordinate) sends a localization request to any of its neighboring anchor nodes. The anchor node sends a reply with its current location reference and its RSSI measurement with respect to the node that wants to localize. Based on this information, we put up a virtual wireless ring (VWR) (or logical ring) [31] as shown in fig. 3. The assumption of the logical ring is made with the anchor node as center. The requesting node can be located anywhere on the circumference of the logical ring, and thus making it difficult to guess its exact location.

b. Next the same requesting node sends another localization request to a different neighboring anchor node. The anchor node follows the same process as discussed in the previous step. Again another logical ring is updated to the previous one, shown in fig. 4. From the logical observation we can analyze that the location of the requesting node could be present in any one of the intersecting point of the two logical rings.

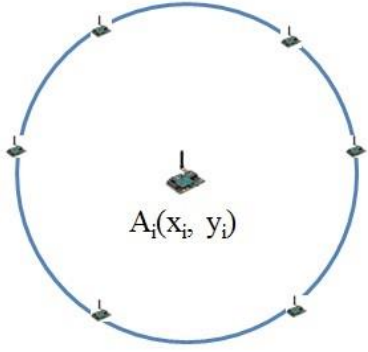
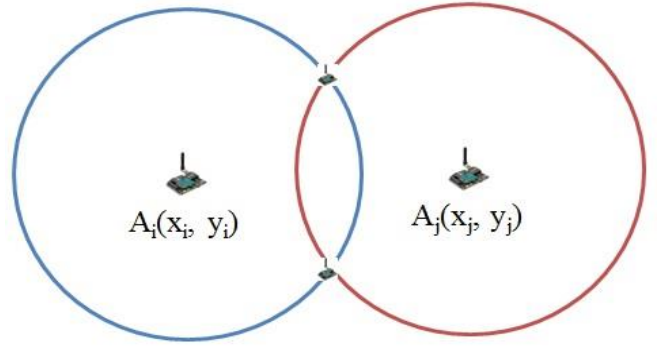

Fig. 3. Virtual wireless ring with one anchor node.        Fig. 4. Virtual wireless ring with two anchor nodes.

c. Finally to ease the muddle, the same requesting node sends another localization request to a different neighboring anchor node other than the previous two anchor nodes. The same process is repeated with the new neighboring anchor node. When the final virtual wireless ring is drawn, we would be able to extract the exact location of the requesting node. Fig. 5 shows the localization of a node using trilateration technique.

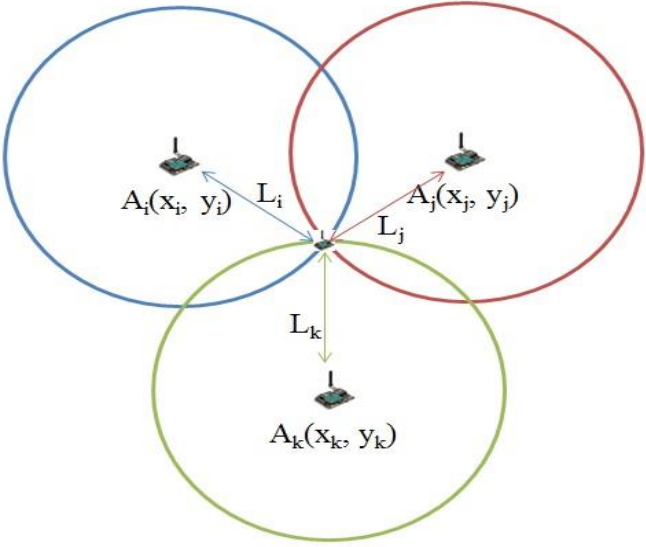
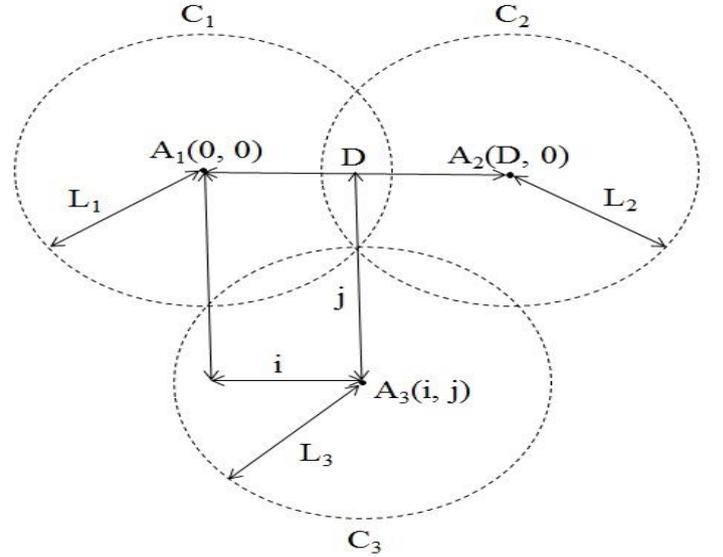

Fig. 5. Virtual wireless ring with three anchor node.        Fig. 6. Trilateration measurements.

The mathematical computation of trilateration is as follows:

Consider three circles or spheres with centre $C_1$, $C_2$ and $C_3$, radius $L_1$, $L_2$ and $L_3$ from points $A_1$, $A_2$ and $A_3$ (anchor node location), refer fig. 6.

The general equation of the sphere is

$$\sum_{k=1}^{3}(A_k - C_k)^2 = L^2$$

The three circles or spheres equation can be modified as follows,

$$L_1^2 = A_1^2 + A_2^2 + A_3^2 \quad (1)$$
$$L_2^2 = (A_1 - D)^2 + A_2^2 + A_3^2 \quad (2)$$
$$L_3^2 = (A_1 - i)^2 + (A_2 - j)^2 + A_3^2 \quad (3)$$

Subtracting equation (2) from equation (1), we get

$$L_2^2 - L_1^2 = (A_1 - D)^2 + A_2^2 + A_3^2 - A_1^2 - A_2^2 - A_3^2 \quad (4)$$

Substituting we get,

$$A_1 = \frac{L_1^2 - L_2^2 + D^2}{2D} \quad (5)$$

From the first two circles ($C_1$ and $C_2$) we can find out that the two circles intersect at two different points, that is

$$D - A_1 < A_2 < D + A_1 \quad (6)$$

Substituting equation (5) in equation (1), we can procure

$$L_1^2 = \left(\frac{L_1^2 - L_2^2 + D^2}{2D}\right)^2 + A_2^2 + A_3^2 \qquad (7)$$

Substituting we get the solution of the intersection of two circles

$$A_2^2 + A_3^2 = L_1^2 - \frac{(L_1^2 - L_2^2 + D^2)^2}{4D^2} \qquad (8)$$

Substituting equation (1) with equation (3), we get

$$L_3^2 = (A_1 - i)^2 + (A_2 - j)^2 + L_1^2 - A_1^2 - A_2^2 \qquad (9)$$

$$A_2 = \frac{L_1^2 - L_2^2 - A_1^2 + (A_1 - i)^2 + j^2}{2j}$$

$$= \frac{L_1^2 - L_2^2 + i^2 + j^2}{2j}$$

$$A_2 = \frac{i}{j} L_1 \qquad (10)$$

From equation (5) and equation (10) we get the values of $B_1$ and $B_2$ respectively. From that we can find out the value of $A_3$ from equation (1),

$$A_3 = \pm\sqrt{L_1^2 - A_1^2 - A_2^2}$$

From the above equation we can say that, $A_3$ can have either positive or negative value. If any circles intersect any other two circles precisely at one point, then $A_3$ will get a value zero. If it intersects at two or more points, outside or inside it can get either a positive or a negative value.

During deployment each node carries out the trilateration process with all of its neighbouring nodes and every node is authorized with two or more trilateration points for security reasons. Every node reveals the information about its trilateration point to its immediate or one hop neighbours. Care is taken that no node reveals the trilateration information about its neighbours.

The algorithm for setting up the anchor nodes according to trilateration are as follows,

```
1.  Start
2.  {
3.  Deploy the anchor nodes
4.  {
5.       Set the initial coordinates (latitude & longitude) for each anchor node
6.       Cluster anchor nodes into a set of three or more
7.  }
8.       Trilaterate a group of anchor nodes to a centre point (or trilateration point) and save the location reference in
         M₁*
9.       Individually trilaterate all the anchor nodes with the neighbouring group and save the location references in
         M₂*, M₃*, etc.
10.      Pass trilateration information to its immediate neighbours
11.      Repeat the above steps for all the anchor nodes
12. }
13. End
(* M₁, M₂, M₃, etc., are different memory with different location reference)
```

The algorithm for finding out the malicious anchor nodes are as follows,

```
1.  Start
2.  {
3.  Trilaterate each group of anchor nodes to a centre point and save that location
4.  {
5.       Compare the obtained location with location reference (M₁)
6.       If comparison not satisfied
7.       {
8.            Trilaterate all anchor nodes (individually) of the particular group (which does not satisfy the above
              comparison) with the neighbouring group (using the trilateration information obtained during
              deployment)
```

```
    9.              Compare the obtained results with the location references (M₂, M₃, etc.)
    10.             {
    11.                 Separate the mismatched anchors node location and save the new location in $M_n$
    12.             }
    13.         }
    14.         If comparison satisfied, no cheating nodes occur
    15. }
    16. End
```

After the comparison, the anchor nodes that does not have the same location reference or the anchor node that tends to be vulnerable is considered to be malicious or cheating node. To confirm its adversary, we compare it with maximum likelihood expectation and Mahalanobis distance.

## 3. Associating with Maximum Likelihood Expectation

One of the most broadly and commonly used classification technique is maximum likelihood expectation / classification. It has a good acceptable result and is extensively employed and demanding algorithm.

The localization error obtained during the above mentioned algorithm is discussed in our next section. And the obtained results are compared with maximum likelihood expectation method. In wireless sensor networks, all the sensor data or the sensed data are sent to a central server or aggregation point. In our scheme, the central server is made available with the location references of all the nodes in the network and MLE method is carried out with the location references available in the aggregation point or the central server.

Maximum likelihood Expectation is a technique that is used in statistics to find the maximum probable value from previously obtained results. The results obtained from maximum likelihood expectation can be used as the parametric values for further experiments or simulations.

### A. Probability density function

Probability density function (pdf) sorts out the required area for the random variable to occur. Consider a random sample ($x_1$, $x_2$, ..., $x_n$) from an unknown population has data vector $x = (x_1, x_2, ..., x_n)$. The probability density function $f(x/w)$ is

$$f(x = (x_1, x_2, ..., x_n)|w) = f_1(x_1|w) * f_2(x_2|w) * ... * f_n(x_n|w)$$

where:
$x$ is a random sample,
$w$ is the parameter value.

Consider a scenario where n (number of trials) = 10, w = 0.4 and x = (0, 1, ..., 10), then the probability density function will be

$$f(x \mid n = 10, w = 0.4) = \frac{10!}{x!\,(10-x)!}(0.4)^x(0.6)^{10-x}$$

The parametric values have a large number of successive probabilities.

### B. Likelihood function

The trilateration groups are denoted as $\varphi_k$, $k = 1, 2, 3, ..., M$ where $M$ is the number of trilateration groups. To determine the group, to which an anchor node with the current location $z$ belongs, the conditional probabilities

$$p(\varphi_k|z), k = 1, 2, 3, ..., M$$

play a crucial role. The probability $p(\varphi_k|z)$ states whether $\varphi_k$ is the correct trilateration group of the anchor node with the give location z. We can categorize the anchor nodes, if we know the complete set of $p(\varphi_k|z)$ from decision rule

$$z \in \varphi_k \text{ if } p(\varphi_k|z) > p(\varphi_n|z) \text{ for all } n \neq k \qquad (11)$$

This explains that the anchor node with location $z$ is the member of group $\varphi_k$ if $p(\varphi_k|z)$ is the largest probability of the set.

The desired $p(\varphi_k|z)$ from the above equation and the available $p(z|\varphi_k)$ from the projected training data, are correlated by Bayes theorem

$$p(\varphi_k|z) = \frac{p(z|\varphi_k)\, p(\varphi_k)}{p(z)} \qquad (12)$$

where:
$(\varphi_k)$ is the probability that anchor nodes from group $\varphi_k$ can move its location,
$p(z)$ is the probability of finalizing an anchor node with location reference $z$.

$$p(\mathbf{z}) = \sum_{k=1}^{M} p(\mathbf{z}|\varphi_k) \, p(\varphi_k)$$

Replacing equation (12) in (11), reduces the decision rule to

$$\mathbf{z} \in \varphi_k \text{ if } p(\mathbf{z}|\varphi_k)p(\varphi_k) > p(\mathbf{z}|\varphi_n)p(\varphi_n) \text{ for all } n \neq k \tag{13}$$

In equation (13), $p(\mathbf{z})$ has been eliminated as a shared factor, since we don't know whether it correct location or false location. As $p(\mathbf{z}|\varphi_k)$ can be obtained from the training data, and it is plausible that the priors $p(\varphi_k)$ can be estimated.

In order to prove mathematically, we define the discriminant function

$$\mathfrak{z}_k(\mathbf{z}) = \ln\{p(\mathbf{z}|\varphi_k)p(\varphi_k)\} = \ln p(\mathbf{z}|\varphi_k) + \ln p(\varphi_k) \tag{14}$$

In order to get a decision rule by substituting equation (14) with equation (13), we need a monotonic function i.e., natural logarithm. We give the decision rule as

$$\mathbf{z} \in \varphi_k \text{ if } \mathfrak{z}_k(\mathbf{z}) > \mathfrak{z}_j(\mathbf{z}) \text{ for all } n \neq k \tag{15}$$

To further proceed with maximum likelihood estimation, a certain probability model is chosen for the trilateration group function $p(\mathbf{z}|\varphi_k)$. In our scheme, we used Gaussian distribution which is as follows:

$$p(\mathbf{z}|\varphi_k) = (2\pi)^{-S/2} |C_i|^{-1/2} e^{-1/2 (\mathbf{z} - \bar{x}_i)^T C_i^{-1} (\mathbf{z} - \bar{x}_i)} \tag{16}$$

where:
$\bar{x}_i$ is the mean position of the anchor node among the trilateration group $\varphi_k$,
$C_i$ is the covariance matrix of the trilateration group $\varphi_k$,
$S$ is the 'N' dimensional space.

To obtain the categorization function, substitute equation (16) with (14)

$$\mathfrak{z}_k(\mathbf{z}) = \frac{-1}{2} S \ln 2\pi - \frac{1}{2} \ln|C_i| - \frac{1}{2} (\mathbf{z} - \bar{x}_i)^T C_i^{-1} (\mathbf{z} - \bar{x}_i) + \ln p(\varphi_k)$$

Simplifying we get,

$$\mathfrak{z}_k(\mathbf{z}) = \ln p(\varphi_k) - \frac{1}{2} \ln|C_i| - \frac{1}{2} (\mathbf{z} - \bar{x}_i)^T C_i^{-1} (\mathbf{z} - \bar{x}_i)$$

Removing the prior probability gives us the trilateration group membership of the anchor node,

$$\mathfrak{z}_k(\mathbf{z}) = -\ln|C_i| - (\mathbf{z} - \bar{x}_i)^T C_i^{-1} (\mathbf{z} - \bar{x}_i) \tag{17}$$

Equation (17) is used to identify whether the anchor node belongs to current trilateration group. If the anchor node does not belong to the group it is considered deceitful. Probability distribution function sorts out the most probable value, which leads for the estimation of expected value.

## 4. Correlating with Mahalanobis Distance

Mahalanobis distance applies posterior probability to identify the outliers. When two anchor nodes in space are demarcated by two or more associated location coordinates, Mahalanobis distance can be used to find the distance measure between the two anchor nodes. Mahalanobis distance identifies the malicious cheating nodes by comparing the location coordinates with respect to a centroid value. In our case the centroid value is the location coordinate of the trilateration point. The Mahalanobis distance function to identify the distance measure between two anchor nodes are as follows:

$$d(mahalanobis) = \left\{ \left[(x_j, y_j) - (x_i, y_i)\right]^T * C^{-1} * \left[(x_j, y_j) - (x_i, y_i)\right] \right\}^{1/2}$$

where:
$d(mahalanobis)$ is the distance between two anchor nodes,
$(x_i, y_i)$ & $(x_j, y_j)$ are the location coordinates of the two anchor nodes,
$C$ is the sample covariance matrix.

The variance-covariance matrix $C$ is constructed in order to gauge Mahalanobis distance,

$$C = \frac{1}{(n-1)} [(x, y)]^T [(x, y)]$$

where:
$(x, y)$ is the matrix containing the location coordinates,
$n$ is the number of nodes.

In the instance of multiple location references the variance-covariance will become as follows:

$$C = \begin{bmatrix} \sigma_1^2(x_i, y_i) & \rho_{12}\,\sigma_1(x_i, y_i)\,\sigma_2(x_j, y_j) \\ \rho_{12}\,\sigma_1(x_i, y_i)\,\sigma_2(x_j, y_j) & \sigma_2^2(x_j, y_j) \end{bmatrix}$$

where:
$\sigma_1^2$ & $\sigma_2^2$ are the variances of the multiple location references,
$\rho_{12}\,\sigma_1(x_i, y_i)\,\sigma_2(x_j, y_j)$ is the covariance between the multiple location references.

The value of $C^{-1}$ is computed as follows:

$$C^{-1} = \begin{bmatrix} \dfrac{\sigma_2^2(x_j, y_i)}{|C|} & \dfrac{-\rho_{12}\,\sigma_1(x_i, y_i)\,\sigma_2(x_j, y_j)}{|C|} \\ \dfrac{-\rho_{12}\,\sigma_1(x_i, y_i)\,\sigma_2(x_j, y_j)}{|C|} & \dfrac{\sigma_1^2(x_j, y_j)}{|C|} \end{bmatrix}$$

where:
$|C|$ is the variance covariance matrix's determinant and is equal to $\sigma_1^2\,\sigma_2^2\,(1 - \rho_{12}^2)$.

The transformation of the location coordinates to matrix form is shown in fig. 7a, b. The Mahalanobis distance function can be modified to identify the distances between multiple location coordinates to a centroid, as follows:

$$d(\delta_i) = \{[(x_i, y_i) - (x_c, y_c)]^T * C^{-1} * [(x_i, y_i) - (x_c, y_c)]\}^{1/2} \text{ for } i = 1, 2, \ldots, n$$

where:
$d(\delta_i)$ is the distance between centroid and $i^{th}$ anchor node,
$(x_i, y_i)$ is the location coordinate of the $i^{th}$ anchor node,
$(x_c, y_c)$ is the location coordinate of the centroid or trilateration point.

The new distances obtained using Mahalanobis distance, are compared using posterior probability; leading to the confirmation of the anchor nodes adversity. The distances obtained using this method is marginally accurate than the previous method.

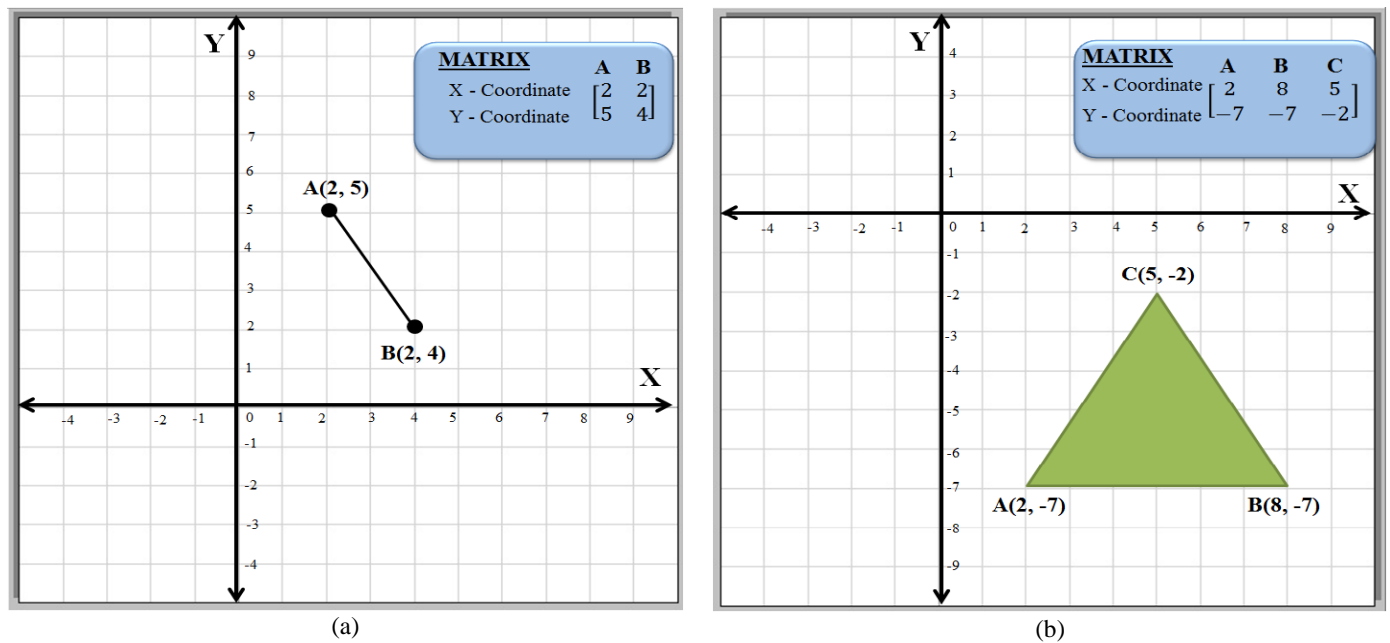

Fig. 7. Transforming location coordinates to matrix representation.

## 5. Simulation and Results

Our simulation was carried out in 600m x 600m two dimensional environment. Deploying the anchor node accurately is very important. First three anchor nodes were placed randomly and the trilateration point is found for the same. An anchor node is placed on the trilateration point attained. Any one of the first three nodes is selected and it acts as the trilateration point of the newer nodes that are going to be deployed. The above process is repeated until the final node is deployed. We deployed around 117 nodes (around 1 node for every 5m x 5m), spread randomly using the above method. Fig. 8 shows the deployment of the anchor nodes in our scenario.

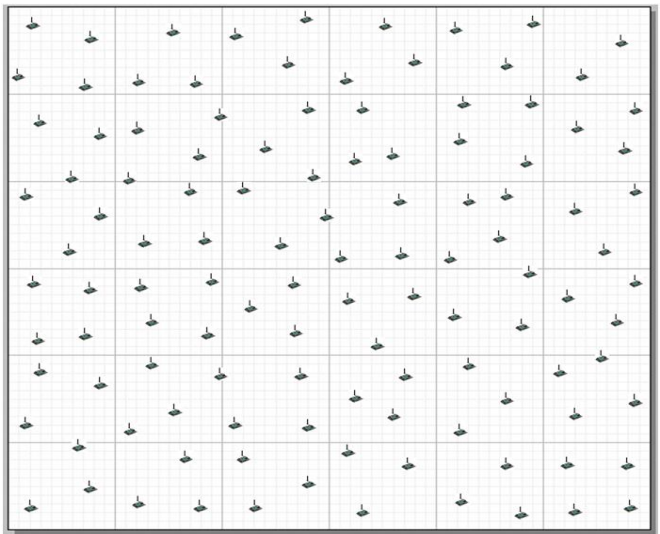

Fig. 8. Deployment of anchor nodes

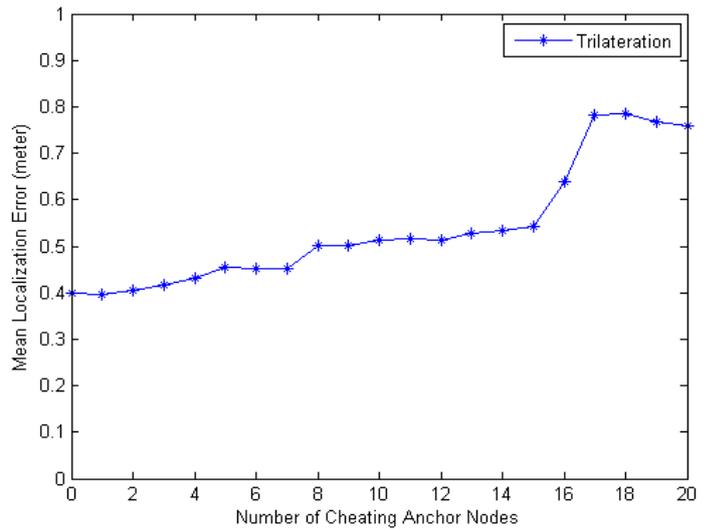

Fig. 9. Mean localization error while using trilateration

### A. Experiment using trilateration technique

Few anchor nodes were compromised (making it transmit false information regarding its current location) randomly and the malicious anchor nodes were found out using trilateration technique. The localization error transpired while localizing the malicious anchor nodes from random samples, were noted down. Each simulation was carried out for 50 times and the mean error was considered. Fig. 9 shows the mean error in location discovery and fig. 10 shows the time taken to locate the malicious anchor nodes during simulation.

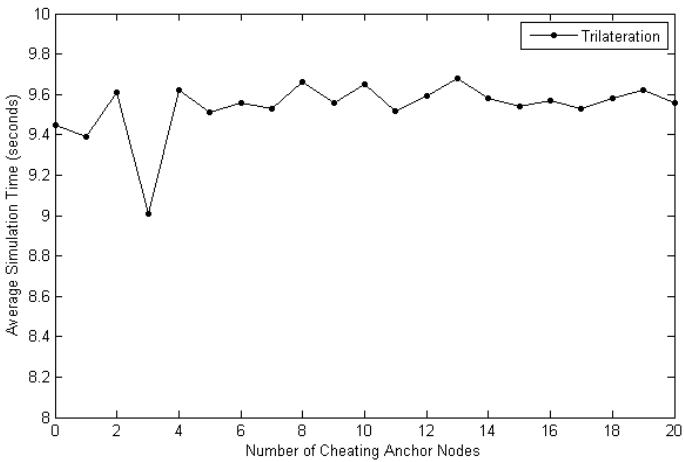

Fig. 10. Average time for simulation.

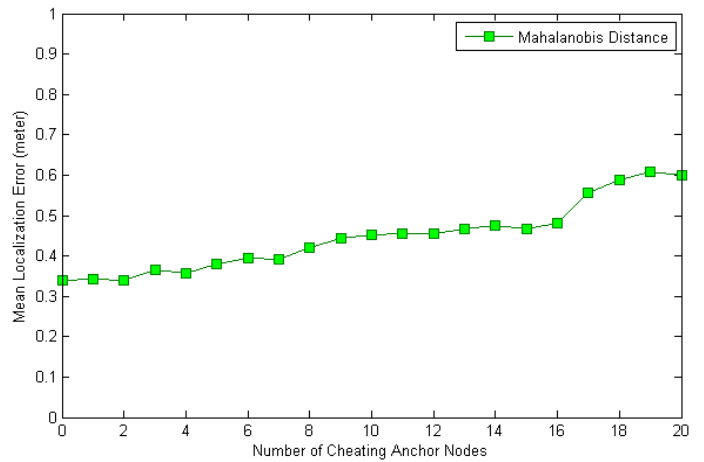

Fig. 11. Mean error after comparing with Mahalanobis distance.

### B. Comparing with Mahalanobis distance

The central server or aggregation point has a list of initial location references of the anchor nodes. The false location of the malicious anchor nodes obtained, were compared with the results obtained from Mahalanobis distance. Comparing the results obtained, reduced the error in location discovery. Fig. 11 shows the mean error in locating malicious anchor nodes.

### C. Comparing with Maximum Likelihood Expectation

Maximum likelihood function has a list of initial location references of the anchor nodes. The false location of the malicious anchor nodes obtained, were compared with the results obtained from maximum likelihood function. Comparing the results obtained, reduced the error in location discovery. Fig. 12 shows the mean error in locating malicious anchor nodes while using maximum likelihood expectation. Fig. 13 shows the comparison of the two results, trilateration and trilateration with MLE.

Finally the information about the malicious anchor node is conveyed to all the nodes other than the infected nodes, and the routing table is updated by confiscating the malicious anchor node.

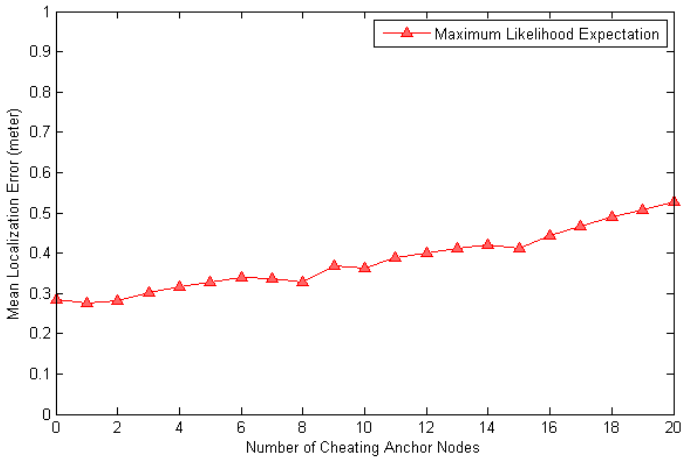

Fig. 12. Mean error after comparing with maximum likelihood function.

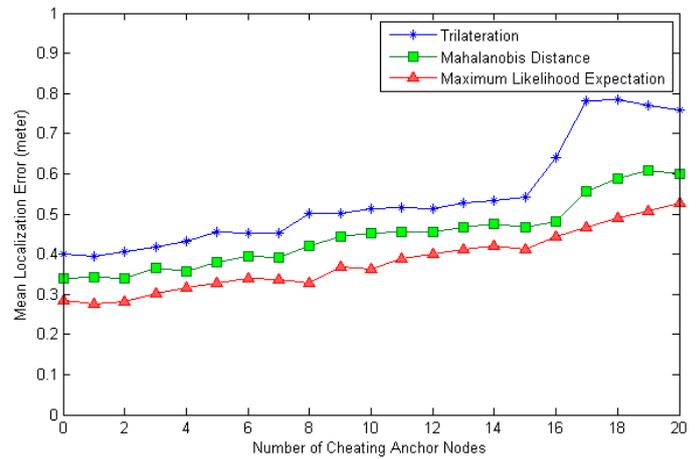

Fig. 13. Result comparison.

## 6. Discussion and Future Events

Malicious anchor nodes will constantly hinder genuine and appropriate localization. Our scheme was carried out using fixed sensor nodes and the attack has a permanent consequence in the sensor node. Reducing the localization error and endorsing the malicious anchor node were implemented successfully in this paper. We have proposed a novel scheme using maximum likelihood and trilateration technique to identify malicious anchor nodes. The error can be increased if hindrance, interferences, and attenuation caused by signal fading, and noise are additional. Our scheme can also be modelled to overcome such disturbances by using some statistical distributions like Rayleigh or Rician distributions [34]. Our algorithm performed consistently for different topologies.

Our scheme can be extended for mobile sensor node with an intermittent attack type. Our framework can be extended to acoustic and ultra-wideband (UWB) technology. Using energy efficiency as a benchmark is quite challenging. Our algorithm was implemented in 2-D plane and can be extended to 3-D plane also.

## 7. Conclusion

For smart environments, security plays a very essential part. In this paper, we discussed about localizing malicious anchor nodes in a secured manner, using trilateration technique and comparing the results obtained with maximum likelihood expectation and Mahalanobis distance. By both the techniques way we were able to reduce the error attained during localization. However, maximum likelihood expectation outperformed Mahalanobis distance in perceiving cheating beacon nodes. By using maximum likelihood expectation and Mahalanobis distance we can obtain consistent and proficient results. Our results show that as the malicious anchor nodes increases, the simulation time and error obtained during location discovery slightly increases. The accuracy obtained in our work can be used as assistance in some wireless applications. Some imminent events for further research have been discussed.